\documentclass{emulateapj}

\newcommand{\deltE}{\Delta\kern-1ptE}

\input epsf

\usepackage{longtable}
\usepackage{amssymb}
\usepackage{graphicx}
\usepackage{epsfig}
\usepackage{epsf}

\slugcomment{To be submitted to ApJ}
\shorttitle{FeK$\alpha$ Fluorescence from II~Peg}
\shortauthors{Ercolano et al.}
\begin{document}
\title{Fe~K$\alpha$ and hydrodynamic loop model diagnostics for a large flare on II~Peg}
\author{Barbara Ercolano$^{1}$, Jeremy J.~Drake$^2$, Fabio Reale$^{3,4}$, 
Paola Testa$^2$ and Jon M.~Miller$^5$}
\affil{$^1$Institute of Astronomy, University of Cambridge,\\
Madingley Rd, Cambridge, CB3 OHA, UK}
\affil{$^2$Harvard-Smithsonian Center for Astrophysics MS-3,\\
60 Garden Street, Cambridge, MA 02138, USA}
\affil{$^3$Dipartimento di Scienze Fisiche ed Astronomiche, 
Sezione di Astronomia, \\ 
Universit`a di Palermo, 
Piazza del Parlamento 1, 
90134 Palermo, Italy}
\affil{$^4$INAF - Osservatorio Astronomico di Palermo, 
Piazza del Parlamento 1, \\
90134 Palermo, Italy}
\affil{$^5$Department of Astronomy, University of Michigan, 
500 Church Street,\\ 
Ann Arbor, MI 48109-1042}

\begin{abstract}
The observation by the {\it Swift} X-ray Telescope of the Fe
K$\alpha_1,\alpha_2$ doublet during a large flare on the RS CVn binary
system II~Peg represents one of only two firm detections to date of
photospheric Fe~K$\alpha$ from a star 
other than our Sun.  We present
models of the Fe~K$\alpha$ equivalent widths reported in the
literature for the II~Peg observations and show that they are most
probably due to fluorescence following inner shell photoionisation of
quasi-neutral Fe by the flare X-rays. Our models constrain the maximum
height of flare the to 0.15~R$_*$ assuming solar abundances for the
photospheric material, and 0.1~R$_*$ and 0.06~R$_*$ assuming depleted
photospheric abundances ([M/H]~=~-0.2 and [M/H]~=~-0.4, respectively).
Accounting for an extended loop geometry has the effect of increasing the
estimated flare heights by a factor of $\sim$3. 
These predictions are consistent with those derived using results of
flaring loop models, which are also used to estimate
the flaring loop properties and energetics. From loop models we
estimate a flare loop height of 0.13~R$_*$, plasma density of
$\sim 4\times10^{12}$~cm$^{-3}$ and emitting volume of
$\sim 6\times10^{30}$~cm$^{3}$. Our estimates for the flare dimensions and
density allow us to estimate the conductive energy losses to
$E_{cond} \leq 2\times10^{36}$~erg, consistent with upper limits
previously obtained in the literature.  Finally, we estimate 
the average energy output of
this large flare to be $\sim10^{33}$~erg~sec$^{-1}$, or
1/10th of the stellar bolometric luminosity.
\end{abstract}


\section{Introduction}\label{s:intro}

\begin{figure*}
\begin{minipage}{17.cm}
\epsscale{0.32}\plotone{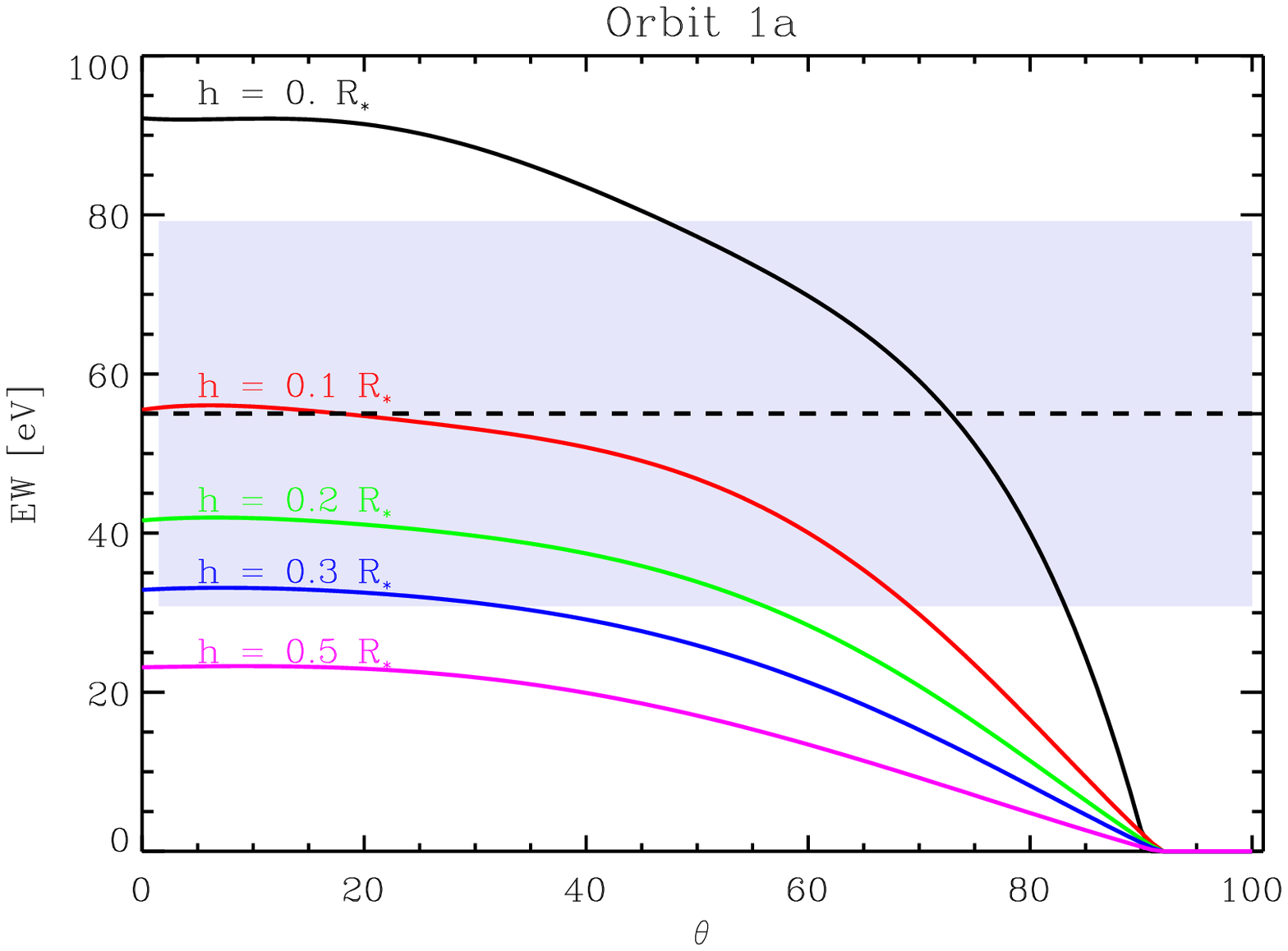}
\epsscale{0.32}\plotone{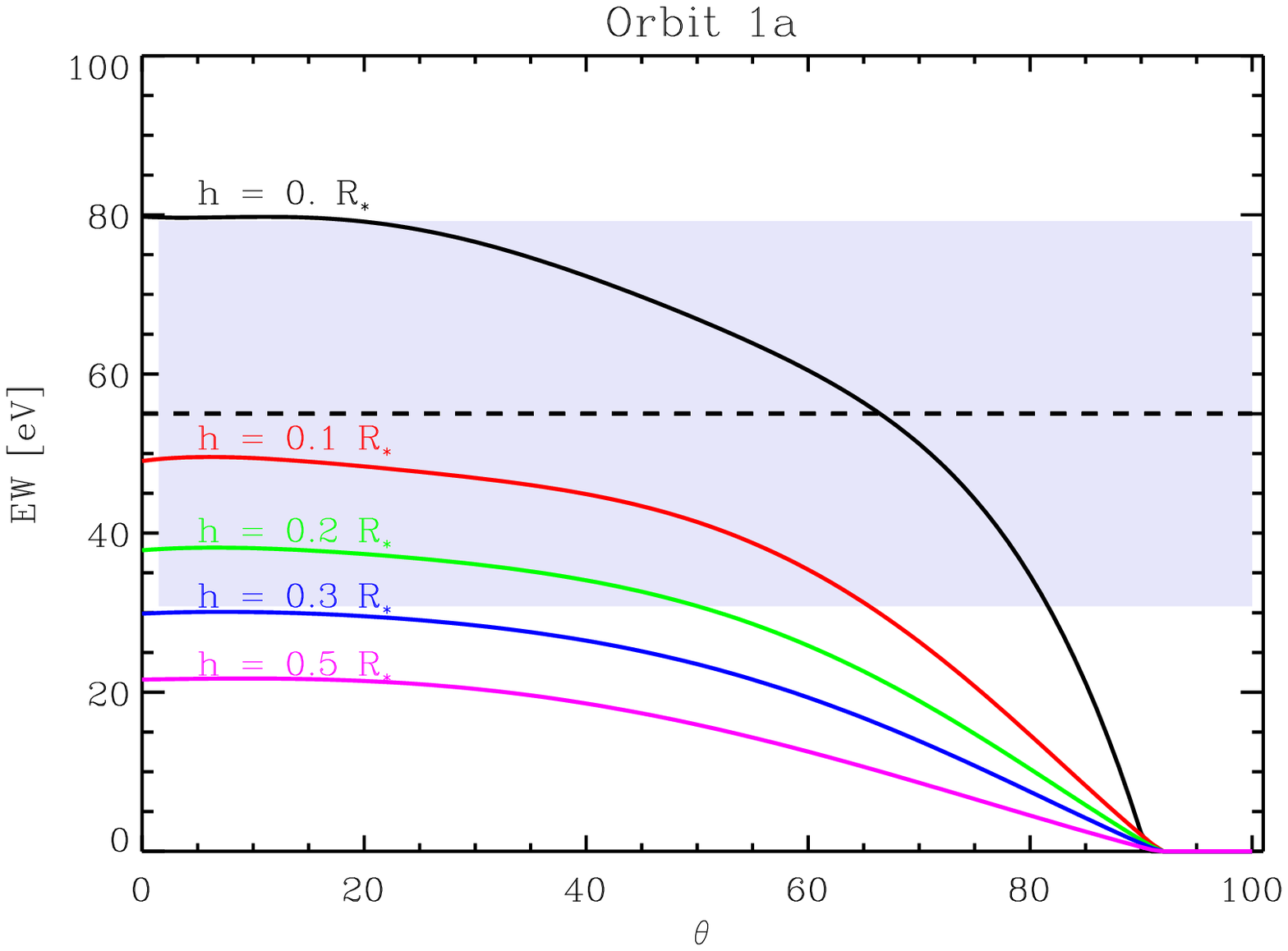}
\epsscale{0.32}\plotone{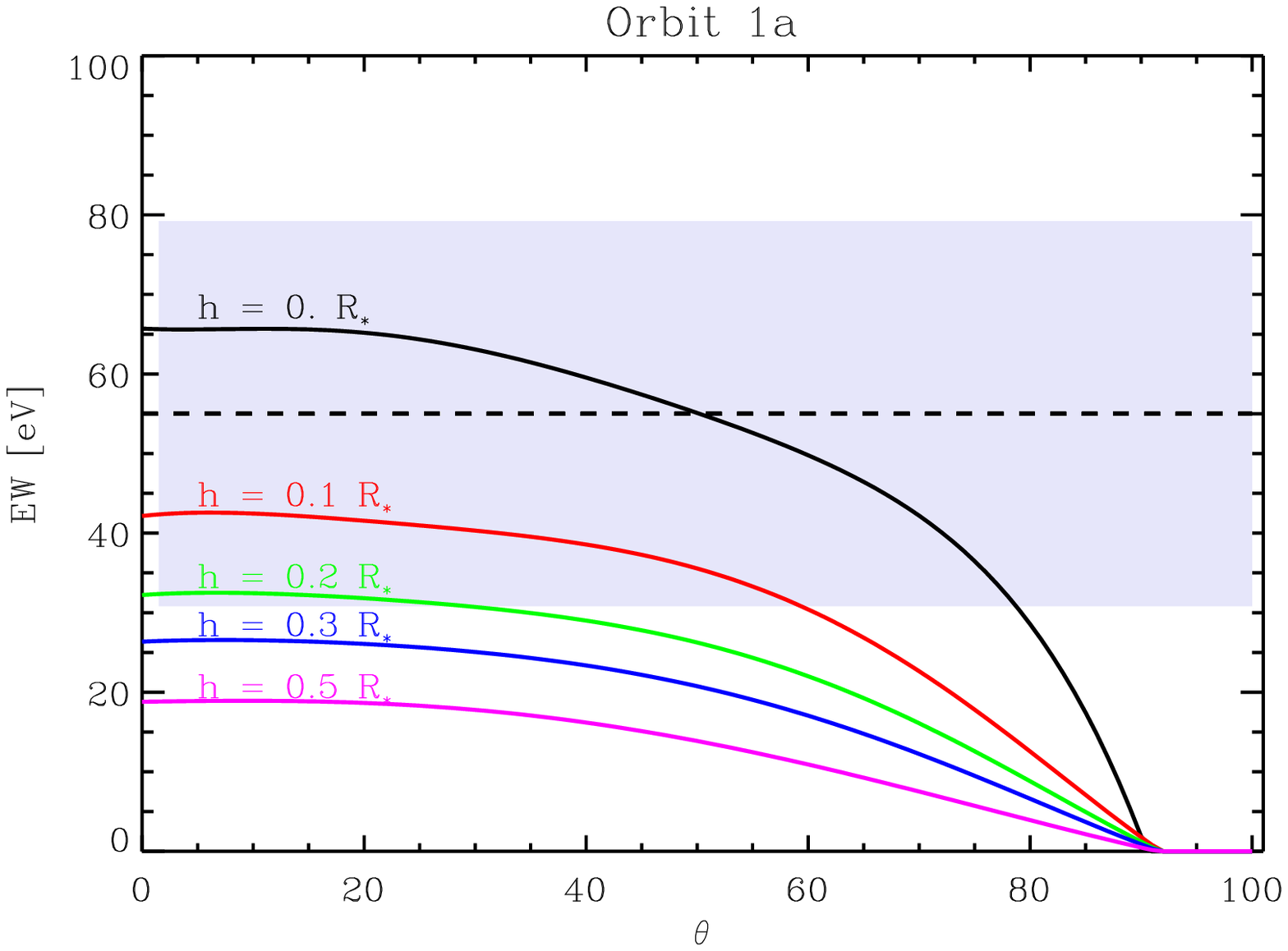}

\end{minipage}
\caption[]{Example Fe~K$\alpha$ equivalent widths as a function of flare
height and inclination angle for three different photospheric metallicities 
for the 1a segment of the flare, as
defined by O07. The observed value and its error is indicated by the horizontal
black dashed line and teh shaded area. Left: Models calculated at solar abundances
(Grevesse \& Sauval, 1998); Central: Models calculated with
[M/H]~=~-0.2; Right: Models calculated with [M/H]~=~-0.4.}
\label{f:f1}
\end{figure*}

A fraction of the X-rays emitted during stellar flares or by hot
stellar coronae are directed downwards towards the stellar
photosphere. Here they interact with the photospheric gas and are
reprocessed through scattering and photoionisation events. The
downward cascade following inner shell photoionisation of
quasi-neutral gas in the stellar photosphere produces characteristic
fluorescent emission from astrophysically abundant species. While the
reprocessed X-ray spectra reach the observer at low flux levels
compared to that of the flares/coronae, fluorescent emission lines,
can be significantly stronger and can be detected against the spectra
of the flare/corona itself.

The intensity or equivalent width of the fluorescent lines detected at
Earth depends uniquely on the details of the fluorescing spectrum, the
photospheric abundance of the photoionised species and the geometry of
the system (flare height and inclination angle; e.g. see fig. 2 of
Testa et al., 2008), making them potentially powerful diagnostics. A number of
theoretical studies exist which provide a framework for the
interpretation of the strong iron fluorescent emission in the solar
context (e.g. Tomblin, 1972; Bai, 1979), and more recently for
arbitrarily photoionised slabs (Kallman et al., 2004) and stellar
photospheres (Drake, Ercolano and Swartz, 2008, DES08). Drake \&
Ercolano (2008a,b) have also recently showed that fluorescent emission
from neon and oxygen should also be detectable in the solar spectrum
with current and future instrumentation. These may provide an
independent measure of the photospheric solar oxygen and neon
abundance, elements that are central to the solution of the ``solar
oxygen crisis'' (e.g. Ayres et al.\ 2006, Socas-Navarro 2007, Basu \&
Antia 2008).

Observationally, the $2s-1p$~6.4~KeV Fe K$\alpha$ doublet from low
ionisation stages of iron has often been detected in solar spectra
(e.g. Neupert et al., 1967; Doschek et al., 1971; Fedelman et al.,
1980; Tanaka et al., 1984; Parmar et al. 1984; Zarro et al., 1992),
from disks around pre-main sequence stars (e.g. Tsujimoto et al.,
2005; Favata et al., 2005), and from the photospheres of the single
G-type giant HR~9024 (Testa et al., 2007) and of the RS CVn binary
system II~Peg (Osten et al., 2007, hereafter O07).

The detection of the 6.4~keV fluorescent iron line in the {\it Swift}
X-ray Telescope spectrum of II~Peg taken during a large flare (O07) is
particularly interesting as it represents the first of only two
detections (with HR~9024, Testa et al., 2007) of photospheric
fluorescence emission in stars other than the Sun. O07, however,
assigned the excitation mechanism to electron impact ionization of
photospheric Fe, rather than photoionisation. In this paper we show by
means of Monte Carlo fluorescence modeling that the Fe~K$\alpha$ 
equivalent widths
reported by O07 are perfectly consistent with photoionisation induced
fluorescence. We also show that our predictions of the flaring loop scale
height agree with those from flaring loop models (Reale, 2007) and
discuss other implications on flaring plasma density, volume and
energetics. 

In Section~2 we briefly describe our Monte Carlo model, and in
Section~3 we present and discuss our results. Section~4 deals with the
flaring loop properties and energetics from hydrodynamic models. Our
final conclusions are given in Section~5.

\section{Monte Carlo Fluorescence Calculations}

The 3D Monte Carlo calculations presented here were performed using a
modified version of the photoionisation and dust radiative transfer
code, MOCASSIN (Ercolano et al. 2003, 2005, 2008). The code uses a a
stochastic approach to the radiation transfer allowing it to deal with
problems of arbitrary geometry, while treating both the primary and
secondary components of the radiation field self-consistently. The
modified version used in this work, which can deal with the production
and transfer of fluorescent radiation, is further described by DES08.

We performed calculations to fit all Fe~K$\alpha$ detections reported
by O07 for the 5 different orbits (flare intervals) of the {\it Swift}
X-ray telescope.  Spectra representing the flaring source of X-ray
irradiation for input into the photospheric fluorescence modeling were
computed for each of these flare intervals.  Spectra were
computed using the CHIANTI database version 5.2 (Dere et al. 1997,
Landi et al. 2006) for the energy range 6.2-30~keV, using the
temperatures and emission measures found by O07 from two-temperature
model fitting and listed in Table~3 of that article.  The upper limit
to the energy range was chosen to be sufficiently high that any
contribution to the observed fluorescent Fe~K line from higher energy
photons in these thermal spectra should be less than 1\%.

Photospheric metal abundances in II~Peg have been studied by Ottmann
et al. (1998) and Berdyugina et al. (1998).  Both are based on high
resolution, high signal-to-noise spectroscopy, and there is little to
choose between the analyses.  The former obtained abundances
[Fe/H]~$=-0.2$, [Mg/H]~$=-0.15$, and [Si/H]~$=-0.15$, expressed relative
to the solar composition in conventional spectroscopic logarithmic
bracket notation, and each with uncertainty of $\pm 0.1$.  As a
verification of their method, they successfully recovered the
currently accepted solar atmospheric parameters from a spectrum of
moonlight.  Berdyugina et al.\ (1998) obtained [M/H]~$=-0.4\pm 0.1$,
and verified their approach against the detailed analysis of the K0
giant Pollux by Drake \& Smith (1991).  It is therefore reasonable to
infer that II~Peg is mildly metal-poor by 0.1-0.4 dex relative to the
Sun.  We have investigated the effects of this slight metal depletion
on Fe~K$\alpha$ equivalent widths by running three sets of models: (i)
with solar abundances (Grevesse \& Sauval, 1998), (ii) with
[M/H]~$=-0.2$ and (iii) with [M/H]$=-0.4$.

Fe~K$\alpha$ equivalent widths also strongly depend on the geometry of
the system, i.e. height of the flare, $h$, and inclination angle,
$\theta$. In particular Fe~K$\alpha$ efficiencies (and therefore
equivalent widths) depend on the function $f(h,\theta)$, which has
been numerically determined for the case of a single flare
illuminating a spherical photosphere by DES08.  Assuming that the
flare can be approximated by a point source above the stellar surface,
we have performed 
calculations at $h~=~0., 0.1, 0.2, 0.3, 0.5$ and $\theta = 0.$ and
used the analytical form of $f(\theta)$ given by DES08 in order to 
estimate the inclination angle dependence.
We note that in order to scale
$\theta = 0$ equivalent widths to an arbitrary value of $\theta$
using the $f(\theta)$ functions one has to first scale the latter such
that $f(0)~=~1$.

\section{Results from Monte Carlo Fluorescence Modeling}

\subsection{Assuming a point source flare}

Representative flare geometry diagnostic diagrams illustrating the
Fe~K$\alpha$ line equivalent width as a function of $\theta$ for
different flare heights are presented in Figure~1 for the three values
of [M/H] (0.0, $-0.2$ and $-0.4$) for Orbit segment 1a.  Also
indicated in these figures for comparison is the observed equivalent
width (dashed line) and uncertainties (shaded area).  
The combination of predicted and observed equivalent widths
can be combined so as to determine the inferred flare height implied
by the observations as a function of the angle $\theta$.  For the grid
of calculations comprising the five different flare segments and three
values of metallicity, we have recast the results into this form in
which the flare point source height implied by the observed equivalent
width is illustrated as a function of $\theta$ in Figure~2.

At solar metallicities, Fe~K$\alpha$
equivalent widths reported for Orbits 1a, 1b, 1c and 2b are well-fit
by models with flare height, $h$, up to $\sim$0.15~R$_*$ for low
($\la$20$^o$) inclination angles; larger values of $\theta$ would
necessarily demand a lower $h$. Lower metallicity models can also
successfully reproduce the observations of Orbits 1a, 1b, 1c and
2b.  Models with [M/H]$=-0.2$ (central column) indicate a maximum flare
height of $\sim0.1$~R$_*$, while models with [M/H]$=-0.4$ indicate a
maximum flare height of $\sim$0.06~R$_*$. We note that the lowest
metallicity models fall slightly short of reproducing the equivalent
width of the Fe~K$\alpha$ line measured during Orbit 2b. The
discrepancy is small enough to be easily explained by observational
uncertainties in the measured equivalent width, which appears
somewhat larger compared with the values determined for the other 
time intervals by O07.  

Finally, the observations during Orbit 2a yield much lower values of
Fe~K$\alpha$ equivalent widths (by a factor of $\sim$3), than for
Orbits 1a, 1b, 1c and 2b. This was already pointed out by O07,
although there was no obvious explanation for the difference.  
According to our model the
Fe~K$\alpha$ equivalent width observed during Orbit 2a implies maximum
flare heights between 0.5 (for [M/H]~=~0) and 0.45~R$_*$ (for
[M/H]~=~-0.4) with an uncertainty of a factor of $\sim 2$.  These 
values are even slightly discrepant with the those deduced
from the other flare intervals.  Intrigued by this discrepancy, we
have independently verified from the {\it Swift} data that the line is
indeed weak during this segment, though we obtain a slightly larger 
95\%\ confidence upper limit of $EW < 43$~eV than found by O07.

\subsection{Effects of extended loop geometry on Fe~K$\alpha$
  fluorescence} 

The above fluorescence calculations and interpretation assume that the
flare can be approximated by a point source of X-rays.  In the case of
large flares similar to those seen on the Sun, with peak temperatures
of order $10^7$~K, the ``hard'' X-ray photons above the 7.11 keV Fe~K
ionisation threshold will be emitted almost uniquely by the hotter
loop apex and the point source approximation should be accurate.
However, Testa et al. (2008) have shown that the extreme flare 
observed on the giant HR~9024 that reached temperatures of $\sim
10^8$~K emits ionising radiation over its entire length, 
such that the equivalent point-source flare height for Fe~K$\alpha$ 
fluorescence was $h/3$, where $h$ is the actual loop height.  

The II~Peg flare analysed here is similar to the HR~9024 one in terms
of plasma temperature, and we can expect a similar scaling of the
point-source height and true flare height.  The corresponding heights
for the [M/H]~$=-0.2$ and $-0.4$ models are $h\sim 0.3$ and
$0.15$~R$_*$, respectively, taking segments 1a, 1b and 1c as
representative.

\section{Flaring loop properties and energetics}

\begin{figure}
\epsscale{0.92}\plotone{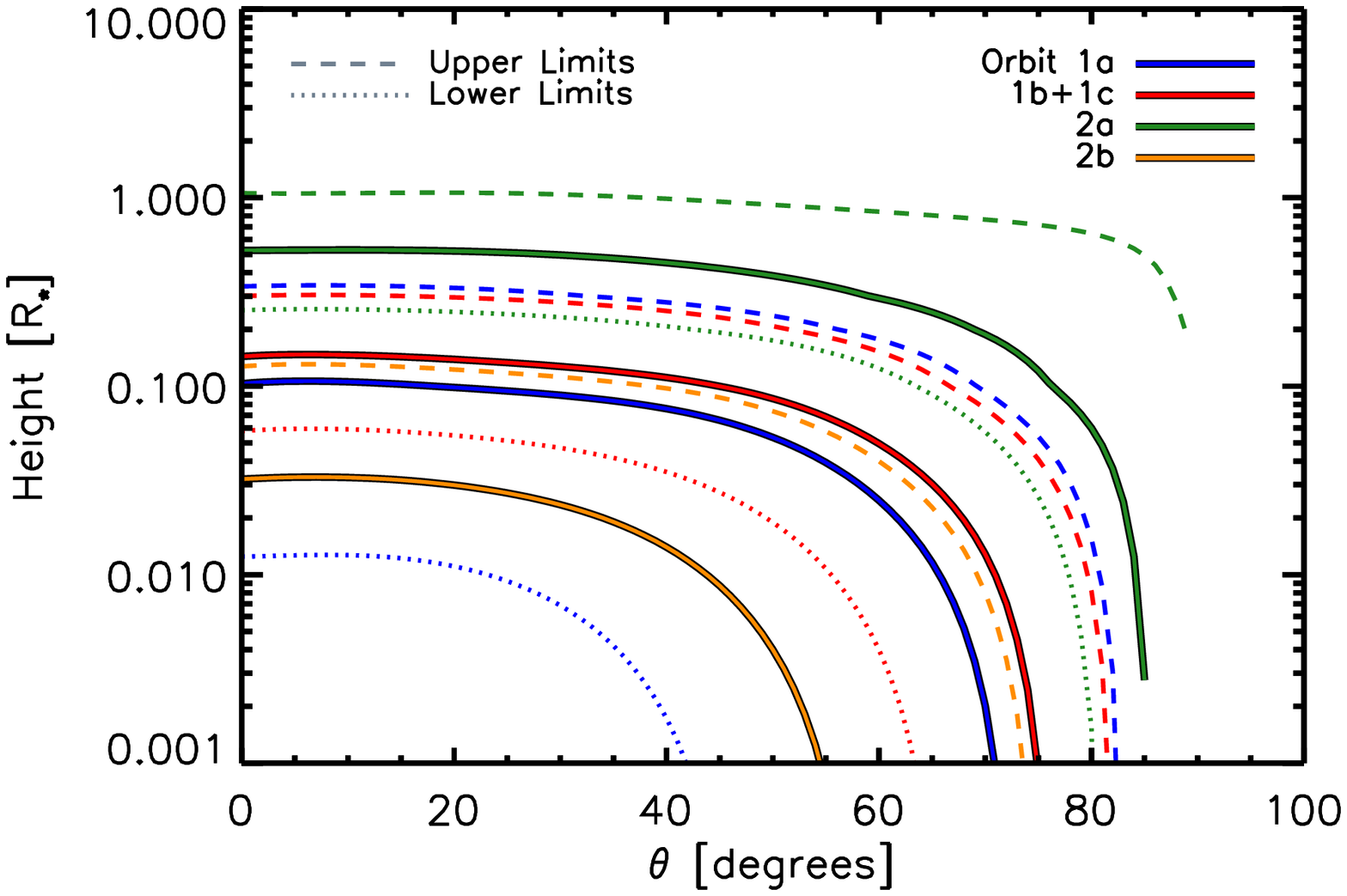} \\
\epsscale{0.92}\plotone{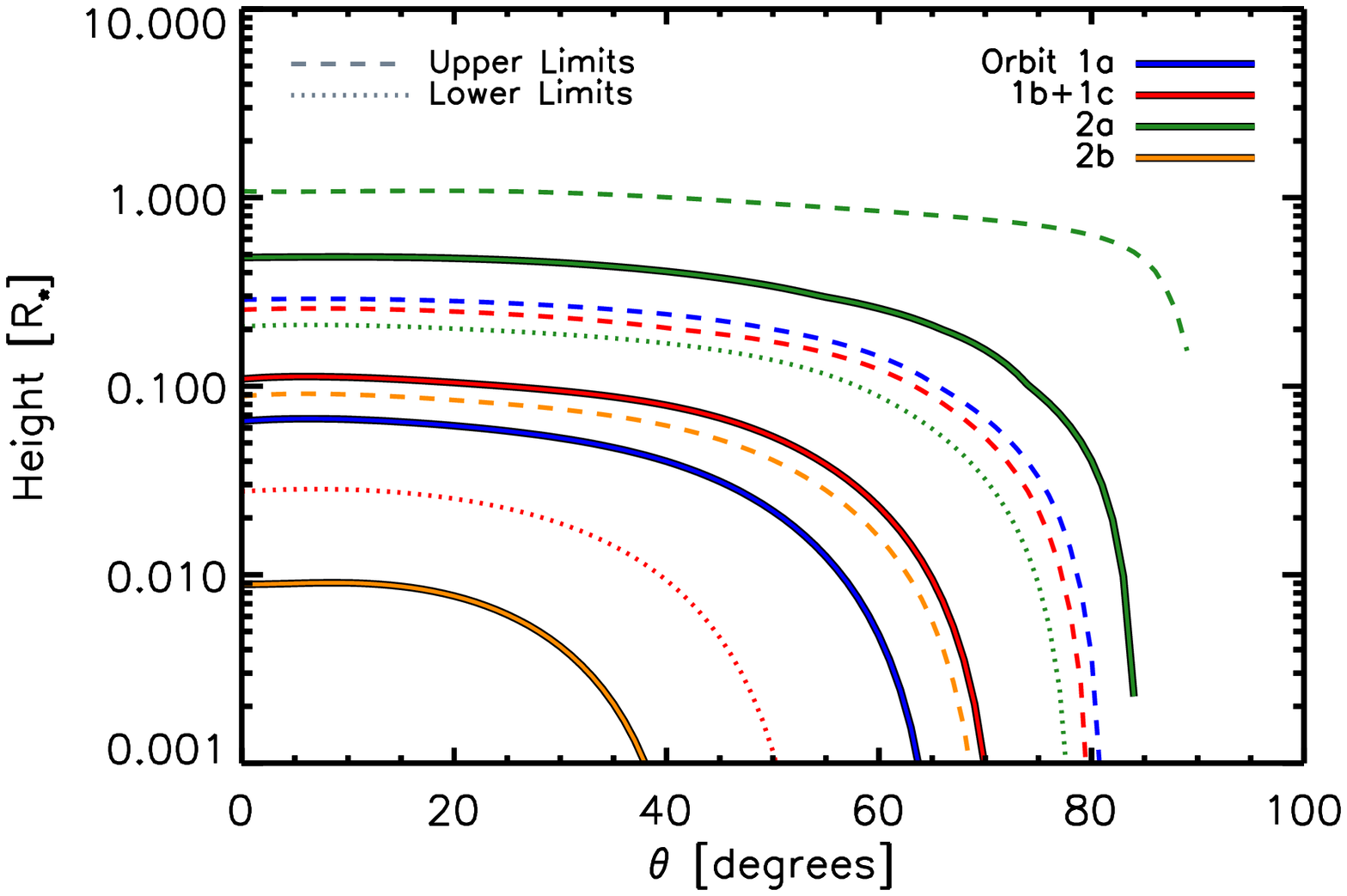} \\
\epsscale{0.92}\plotone{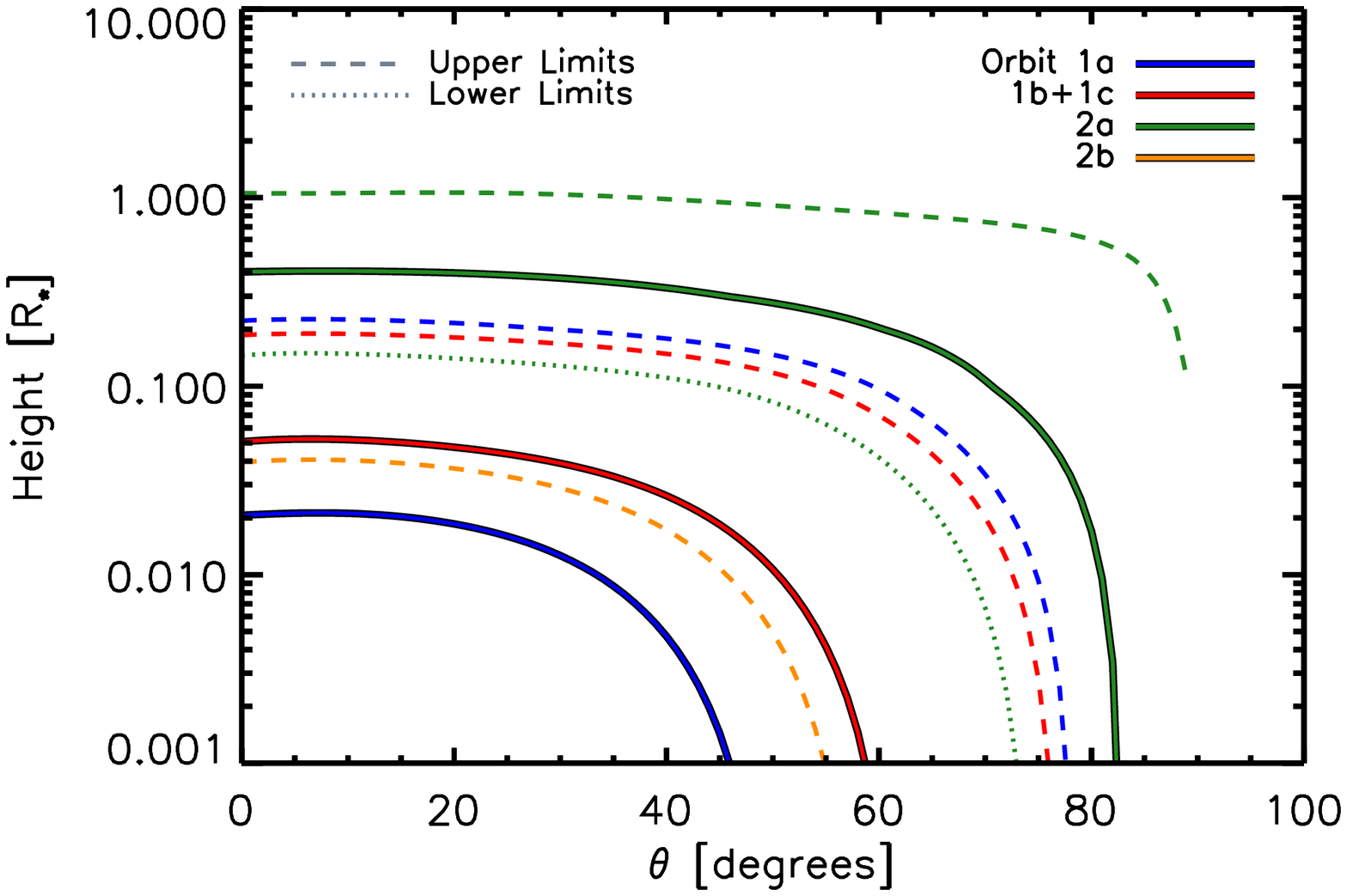}
\caption[]{Inferred flare height vs astrocentric angle for the different
flare segments computed for the different photospheric metallicities 
considered here: [M/H]=0.0 (top); [M/H]$=-0.2$ (middle); and [M/H]$=-0.4$ 
(bottom).  The observed X-ray spectral properties and Fe~K$\alpha$ 
equivalent widths for 
segments 1b and 1c are essentially identical and have been averaged.  
The different segments are colour-coded, with upper and lower limits 
given by dashed and dotted curves, respectively. }
\label{f:h_ang}
\end{figure}

The allowed ranges of flaring loop scale height derived from the
strength of the Fe~K line in the previous section can be compared with
predictions from detailed hydrodynamic flare models.  Here, we use the
scaling laws for loop length, cross-sectional area and plasma density
derived from hydrodynamic simulations by Reale (2007) in application to the
light curve and plasma emission measure and temperature found from
the {\it Swift} observations by O07.  

Relevant quantities required by the Reale (2007) relations are the
flare rise time $t_{M,3}$ (in units of $10^3$s), the maximum plasma
temperature $T_{0,7}$ (in units of $10^7$~K), the peak emission measure
$\Phi$ (cm$^{-3}$), and the $1/e$ decay time $\tau^\prime_s$ (s).
We estimate the flare rise and decay times from Fig.~1 in O07
al.\ (2007), and take the flare temperature and emission measure data
from their Table 2.  Since we are interested in the peak flare
temperature, we adopt the highest temperature obtained from their
3-temperature best-fit models.  The final set of adopted parameters is
\begin{eqnarray}
t_{M,3} & \approx & 2 \\ 
T_{0,7} & \approx & 30 \\
\Phi & \approx & 10^{56} \\
\tau^\prime_s & \approx & 9000 
\end{eqnarray}
Here, we have assumed that the maximum plasma temperature is the value
observed at the time of maximum density, and that this time corresponds to the
peak in the observed flare emission.  Since the flaring loop is a
closed volume, the latter assumption is straightforward. However, as noted by
Reale (2007), the maximum flare temperature can be reached some time after 
the peak density is reached; we allow for this in the analysis below. 

\subsection{Loop length}

The flaring loop half-length can be estimated from the from rise time 
and the maximum plasma temperature, as given in Eqn.~(12) of Reale (2007):
\begin{equation}
L_9 \approx 3 ~ \psi^2 T_{0,7}^{1/2} t_{M,3}
\label{eq:lris}
\end{equation}
where $L_9$ is the loop half-length in units of $10^9$~cm.  The
parameter $\psi$ is the ratio of the maximum plasma temperature to that at 
density maximum,
\[
\psi = \frac{T_0}{T_{M}}.
\]
From Fig.~2 of O07, it can be seen that the rise
phase flattens toward the emission maximum; this suggests the flaring
loop is close to equilibrium at the maximum and that, therefore, $\psi
\approx 1$ and the assumption of maximum plasma temperature at the
time of maximum density holds.  To allow for deviations from this
assumption, we also quote results for $\psi \approx 1.3$ (see examples
given in Reale 2007).  For $T_{0,7} \approx 30$, $t_{M,3} \approx 2$,
$\psi \approx 1$
\[
L_9 \approx 30
\]
For $\psi \approx 1.3$

\[
L_9 \approx 70
\]
As a reasonable average value, we adopt $L_9 \approx 50 \pm 20$.
Adopting a radius for II~Peg of $3.4R_\odot$ (Berdyugina et al.\
1998), this loop half-length corresponds to a loop {\em height} of
$\sim0.13R_\star$.  This compares very favourably with the values we have
estimated from the Fe~K fluorescence line in the previous section,
provided the flare were located toward the centre of the visible
stellar disk and not toward the limb.

As a verification that this loop length value is reasonable, we can
also obtain an estimate of the {\em upper limit} to $L_9$
from the decay timescale.  If the decay is entirely driven by cooling,
with no further heat input, the observed decay timescale can be
equated with the thermodynamic decay time,
\begin{equation}
\tau^\prime_s = \phi \tau_s
\label{eq:taunew}
\end{equation}
where $\phi \approx 1.3$ and 
\begin{equation}
\tau_{s} = 3.7 \times 10^{-4} \frac{L}{\sqrt{T_0}}
= 120 \frac{L_9}{\sqrt{T_{0,7}}}
\label{eq:tserio}
\end{equation}
For $\tau^\prime_s \approx 9000$~s, we obtain an upper limit for
the loop length:
\[
L_9 \leq 320. 
\]
While much larger, this upper limit is 
consistent with the value obtained from the rise phase above.

Huenemoerder et al.\ (2001) observed a large flare on II~Peg during a
45~ks observation with the {\it Chandra} High Energy Transmission
Grating Spectrometer (HETGS), during which the count rate was observed
to rise to 2.5 times its quiescent value and the peak flare
temperature reached $\log T \sim 7.6$.  While much less intense than
the {\it Swift} flare discussed here, the decay timescale of $\sim
65$~ks was somewhat longer.  Huendmoerder et al.\ (2001) interpreted
the light curve in terms of both single loop and two-ribbon type
flares and inferred a loop height in the range 0.05--0.25~R$_\star$
for the former, depending on the plasma density assumed.  This height
then scales inversely with the cube of the number of loops
involved---e.g.\ smaller by a factor of ~5 for 100 loop strands.
While this estimate must be considered very approximate, the
similarity with the loop height derived here for the {\it Swift} flare
suggests that both events occurred in similar coronal structures that
might be typical of the outer atmosphere of II~Peg.

\subsection{Plasma Density, Volume and Loop Cross-sectional Area}

From Reale (2007), the maximum possible density equilibrium value at
the loop apex in units of 
$10^{10}$~cm$^{-3}$ is given by 
\[
n_{0,10} = 13 \frac{T_{0,7}^2}{ L_9}
\]
While this is an upper limit, the true value should not be lower by 
more than a factor of 2.  For $T_{0,7} \approx 30$ and $L_9 \approx 50$, the 
density is then 
\[
n_{0,10} \leq 360
\]

A reasonable value to adopt is $n_{0,10} \approx 200$. This value is compatible with density diagnostics from line ratios of
hot He-like triplets, such as Mg~{\sc xi}, observed during flaring and
quiescence on II~Peg (Huenemoerder et al.\ 2001; Testa, Drake \& Peres
2004). Similarly high densities have also been reported by G{\"u}del et al. (2002) during an X-ray flare in 
Proxima Centauri.
From Reale (2007), the volume is:
\begin{equation}
V \approx \frac{\Phi}{n_{avg}^2}
\label{eq:vol}
\end{equation}
where 
\begin{equation}
n_{avg} = n_M \frac{T_M}{T_{avg}}
\label{eq:navg}
\end{equation}
Taking $T_{avg} \approx 150$ MK from the {\em single temperature} model 
fit of O07 (their Table 2), we obtain
\[
n_{avg} \approx 4 \times 10^{12} ~~~~~ \rm cm^{-3}
\]
and 
\[
V \approx 6 \times 10^{30} ~~~~~ \rm cm^3
\]

The cross-sectional area is:
\begin{equation}
A \approx \frac{V}{2 ~ L}
\label{eq:area}
\end{equation}
For $L_9 \approx 50$, $A \approx 6 \times 10^{19}$ cm$^2$. For a
single loop with circular cross-section we obtain a radius in units of
$10^9$~cm of $r_9 \approx 4$, which is slightly less than 1/10 the
loop half-length and very similar to the values found for solar
flaring loops (e.g. Cheng et al.\ 1980, Golub et al.\ 1980, Peres et al.\ 
1987).

\subsection{Conductive energy losses}

One of the most striking aspects of the {\it Swift} II~Peg flare is
the potentially large flare energy budget.  Using the classical
Spitzer formulism, O07 found an upper limit to the conductive
losses of $5\times 10^{43}$~erg, but were unable to constrain this
further without estimates of the flare dimensions and density.  The
bolometric luminosity of II~Peg is $\log L_{bol}=34.2$ (Marino et al.\
1999), and conductive losses as high as $10^{43}$~erg during a single,
compact flare would represent an astounding concentration of energy.

We can refine the estimate of the conductive energy losses
using Eqn.~(5) in O07 and the parameters derived above
for the flaring loop:
\[
E_{cond} = \frac{\kappa T^{7/2} \Phi \Delta t}{L^2 n_e^2},
\]
where, for Orbit 1, $T \approx 300$ MK (an upper limit for the flare as
a whole), $\Phi \approx 10^{56}$, $\kappa \approx 10^{-6}$ (in
c.g.s. units), $\Delta t \approx 2$ ks, $L_9 \approx 50$, $n_e \approx
n_{avg} \approx 4 \times 10^{12}$ cm$^{-3}$, we obtain
\[
E_{cond} \leq 2 \times 10^{36} ~~~ \rm erg
\]
which is similar to the total radiated energy of $6\times 10^{36}$~erg
in the 0.01-200 keV energy band estimated by O07.
The duration of the flare was of order $10^4$s, and the radiative and
conductive losses therefore represent an energy output of $\sim
10^{33}$~erg~s$^{-1}$, or about 1/10 the stellar bolometric
luminosity.

\section{Conclusions}

We have presented a set of Monte Carlo calculations able to reproduce
the Fe~K$\alpha$ equivalent widths from the {\it Swift} X-ray
Telescope observations of II~Peg during a 'superflare' (Osten et al.,
2007, O07). Our models show that the data are consistent with the
Fe~K$\alpha$ emission being produced by fluorescence following K-shell
photoionisation of quasi-neutral iron in the stellar photosphere.
This contrasts with the interpretation of O07, who favoured
collisional ionisation by non-thermal electrons to produce the
Fe~K$\alpha$ emission. They argued that the normal incidence flare
X-ray penetration depth required in the photosphere to obtain the
observed equivalent widths is similar to the $\tau=1$ Compton
scattering depth, and consequently any fluorescent photons produced
would not easily escape.  However, this assessment was based on a
simple analytical formula appropriate for optically thin cases in
which only a small fraction of the incident X-ray flux undergoes
photoabsorption or scattering (e.g. Liedahl, 1999; Krolik \& Kallman,
1987).  The semi-infinite photospheric case lies outside the range of
applicability of this formula.  Furthermore, incident angles on the
photosphere range from $\sim 0$--$90^\circ$ and path lengths for
escape can therefore be much smaller than gas penetration depths (by a
factor equal to the inverse cosine of these angles).  Our stochastic
treatment includes Compton scattering and shows that the observed
equivalent widths can be produced by fluorescence.  We also note that
the impact excitation mechanism is a very low efficiency process and
requires a large amount of energy in the form of accelerated
electrons.  This was already noted by, e.g., Parmar (1984) in
reference to solar Fe fluorescence and by Ballantyne (2003), and was
discussed in reference to a large flare on II~Peg by Testa et
al. (2008).

The derived flare loop heights are $h$~=~0.15~R$_*$, assuming solar
photospheric abundances, $h$~=~0.1~R$_*$ assuming photospheric
abundances depleted by [M/H]~=~-0.2 and $h$~=~0.06~R$_*$ assuming
[M/H]~=~-0.4. These value are in good agreement with the predictions
from our alternative analysis based on hydrodynamic models
(e.g. Reale, 2007) which yield $h \approx 0.13$~R$_*$. We estimate
flaring loop properties and energetics using scaling laws based on the
hydrodynamic loop modeling of Reale (2007) and obtain a plasma
density of $4\times10^{12}$~cm$^{-3}$ and volume of
$6\times10^{30}$~cm$^{3}$. Using the derived values for the flare
dimensions and densities we obtain a more stringent upper limit for
the conductive energy loss, $E_{cond}$, than previously possible,
setting $E_{cond} \leq 2\times10^{36}$~erg. Considering the duration
of the flare and the radiative and conductive losses, the energy
output of the flare is estimated at $\sim$10$^{33}$~erg sec$^{-1}$,
which is approximately one tenth of the stellar bolometric luminosity.

While we cannot rule out a contribution to the observed Fe~K$\alpha$
flux from electron impact with nonthermal electrons, as proposed by
O07, such contribution is not necessary to explain the available data.

\section*{Acknowledgments}

We thank the anonymous referee and the editor Eric Feigelson for helpful comments that added to the clarity of the paper and the interpretation of the results. JJD was supported by the Chandra X-ray Center NASA contract NAS8-39073 during the course of this research. The simulations were run on the Cosmos 
(SGI altix 4700) supercomputer at DAMTP in Cambridge. Cosmos is a 
UK-CCC facility which is supported by HEFCE and STFC.


\subsection*{References}

\begin{description}
\item[] Ayres, T.~R., Plymate, 
C., \& Keller, C.~U.\ 2006, \apjs, 165, 618 

\item[] Basu, S., \& Antia, H.~M.\ 2008, \physrep, 457, 217 

\item[] Socas-Navarro, H., \& Norton, A.~A.\ 2007, \apjl, 660, L153 

\item[] Berdyugina, S.~V., Jankov, S., Ilyin, I., Tuominen, I., \& 
  Fekel, F. C. 1998, A\&A, 334, 863

\item[] Cheng, C.-C.,
Tandberg-Hanssen, E., \& Smith, J.~B., Jr.\ 1980, Sol.Phys., 67, 259

\item[] Dere, K.~P., Landi, E., Mason, H.~E., Monsignori Fossi, B.~C., 
\& Young, P.~R.\ 1997, A\&AS, 125, 149

\bibitem[Drake et al.(2008)]{2008ApJ...678..385D} Drake, J.~J., Ercolano, 
B., \& Swartz, D.~A.\ 2008, \apj, 678, 385, DES08

\bibitem[Drake 
\& Ercolano(2007)]{2007ApJ...665L.175D} Drake, J.~J., \& Ercolano, B.\ 2007,
\apjl, 665, L175 

\bibitem[Drake 
\& Smith(1991)]{1991MNRAS.250...89D} Drake, J.~J., \& Smith, G.\ 1991, \mnras,
250, 89 

\item[] Ercolano, B.,
Barlow, M.~J., Storey, P.~J., \& Liu, X.-W.\ 2003a, \mnras, 340, 1136 

\item[] Ercolano, B., Young, 
P.~R., Drake, J.~J., 
\& Raymond, J.~C.\ 2007, ArXiv e-prints, 710, arXiv:0710.2103 

\item[] Golub, L., Maxson, C., 
Rosner, R., Vaiana, G.~S., \& Serio, S.\ 1980, ApJ, 238, 343

\item[] G{\"u}del, M., 
Audard, M., Skinner, S.~L., \& Horvath, M.~I.\ 2002, \apjl, 580, L73

\item[] Huenemoerder, 
D.~P., Canizares, C.~R., \& Schulz, N.~S.\ 2001, \apj, 559, 1135 

\item[] Krolik, J.~H., \& Kallman, T.~R.\ 1987, \apjl, 320, L5 

\item[] Landi, E., Del Zanna, G., 
Young, P.~R., Dere, K.~P., Mason, H.~E., \& Landini, M.\ 2006, ApJS, 162, 261 

\item[] Liedahl, D.~A.\ 1999, X-Ray Spectroscopy in Astrophysics, 520, 189 

\item[] Marino, G., Rodon\'o, M., Leto, G., \& Cutispoto, G. 1999, A\&A, 
352, 189

\item[] Osten, R.~A., Drake, S., Tueller, J., Cummings, J., Perri, M.,
Moretti, A., \& Covino, S.\ 2007, ApJ, 654, 1052

\item[] Peres, G., Reale, F., 
Serio, S., \& Pallavicini, R.\ 1987, ApJ, 312, 895 

\item[] Reale, F.\ 2007, A\&A, 471, 271

\bibitem[Testa et al.(2004)]{2004ApJ...617..508T} Testa, P., Drake, J.~J., 
\& Peres, G.\ 2004, \apj, 617, 508 

\item[] Testa, P., Drake, J.~J., Ercolano, B., Reale, F., Huenemoerder, D.~P., Affer, L., Micela, G., \& Garcia-Alvarez, D.\ 2008, \apjl, 675, L97

\end{description}

\end{document}